\documentstyle[12pt]{article}
\renewcommand{\theequation}
{\arabic{section}.\arabic{equation}}

      \newcommand{\beq}{\begin{equation}}
      \newcommand{\eeq}{\end{equation}}
      \newcommand{\beqa}{\begin{eqnarray}}
      \newcommand{\eeqa}{\end{eqnarray}}
      
      \newcommand{\nn}{\nonumber}
      \newcommand{\del}{\partial}
      \newcommand{\ddel}{{\stackrel{\leftrightarrow}{\del}}}
      \newcommand{\dlangle}{\left\langle\left\langle}
      \newcommand{\drangle}{\right\rangle\right\rangle}
      \newcommand{\al}{\alpha}
      \newcommand{\be}{\beta}
      \newcommand{\ga}{\gamma}
      \newcommand{\de}{\delta}
      
      \newcommand{\th}{\theta}
      \newcommand{\ka}{\kappa}
      \newcommand{\la}{\lambda}

      \newcommand{\si}{\sigma}
      \newcommand{\phc}{{\phi}_c}
      \newcommand{\cb}{{\bar{c}}}
      \newcommand{\df}{{\sqrt{-g}}}
      \newcommand{\ind}[2]{\int d^{#1}{#2}}
      \newcommand{\gF}{{g_F}}
      \newcommand{\ap}{{a^{\prime}}}
      \newcommand{\kp}{{k^{\prime}}}
      
      \newcommand{\kpp}{{k^{\prime\prime}}}
      
      \newcommand{\brs}{{\mbox{\boldmath$\delta$}_B}}
      \newcommand{\Vi}{{{\cal V}_{in}}}
      \newcommand{\Vo}{{{\cal V}_{out}}}
      \newcommand{\lag}{{\cal L}}
      \newcommand{\tilag}[1]{{\widetilde{\lag}_{#1}}}
      \newcommand{\bolU}{{\mbox{\boldmath$U$}}}
      \newcommand{\bolUdag}{{{\mbox{\boldmath$U$}}^{\dag}}}
      \newcommand{\bolrh}{{\mbox{\boldmath$\rho$}}}
      \newcommand{\bolrhdag}{{{\mbox{\boldmath$\rho$}}^{\dag}}}
      \newcommand{\bolB}{{\mbox{\boldmath$B$}}}
      \newcommand{\bolBdag}{{{\mbox{\boldmath$B$}}^{\dag}}}
      \newcommand{\bolpi}{{\mbox{\boldmath$\pi$}}}
      \newcommand{\bolpidag}{{{\mbox{\boldmath$\pi$}}^{\dag}}}
      \newcommand{\bolc}{{\mbox{\boldmath$c$}}}
      \newcommand{\bolcdag}{{{\mbox{\boldmath$c$}}^{\dag}}}
      \newcommand{\bolcb}{{\bar{\mbox{\boldmath$c$}}}}
      \newcommand{\bolcbdag}{{{\bar{\mbox{\boldmath$c$}}}^{\dag}}}
      \newcommand{\bolQB}{{\mbox{\boldmath$Q$}_B}}
      \newcommand{\bolQc}{{\mbox{\boldmath$Q$}_c}}
      
      \newcommand{\brB}{{\breve{B}}}
      \newcommand{\brc}{{\breve{c}}}
      \newcommand{\brcb}{{\breve{\cb}}}
      \newcommand{\EM}[1]{{T_{{#1}\:\mu\nu}}}
      \newcommand{\alU}{{\al_{U}}}
      \newcommand{\beU}{{\be_{U}}}
      \newcommand{\alBp}{{\al_{B\,\pi}}}
      \newcommand{\beBp}{{\be_{B\,\pi}}}
      \newcommand{\balBp}{{\bar{\al}_{B\,\pi}}}
      \newcommand{\bbeBp}{{\bar{\be}_{B\,\pi}}}
      \newcommand{\alcc}{{\al_{\cb\,c}}}
      \newcommand{\becc}{{\be_{\cb\,c}}}
      \newcommand{\balcc}{{\bar{\al}_{\cb\,c}}}
      \newcommand{\bbecc}{{\bar{\be}_{\cb\,c}}}
      \newcommand{\alr}{{\al_{\rho}}}
      \newcommand{\ber}{{\be_{\rho}}}
      \newcommand{\tilI}{{\widetilde{I}}}

\begin{document}
\begin{titlepage}
    \begin{normalsize}
     \begin{flushright}
                 UT-Komaba/93-9 \\
                 May 1993
     \end{flushright}
    \end{normalsize}
    \begin{LARGE}
       \vspace{1cm}
       \begin{center}
         {Cancellation of  unphysical gauge and ghost\\
          degrees of freedom in backreaction} \\
       \end{center}
   \end{LARGE}

  \vspace{5mm}

\begin{center}
           Masaru O{\sc noda}
           \footnote{E-mail address:
              onoda@tkyvax.phys.s.u-tokyo.ac.jp}\\
      \vspace{4mm}
						{\it Institute of Physics, College of Arts and Sciences} \\
	     {\it University of Tokyo, Komaba}\\
      {\it Meguro-ku, Tokyo 153, Japan}\\
      \vspace{1cm}

    \begin{large} ABSTRACT \end{large}
\par
  \end{center}
\begin{quote}
  \begin{normalsize}
\ \ \ \
     We study the U(1) Higgs model in spacetime-dependent background
     fields (a background metric and a background scalar field).
     Particle creation can occur because of the time-dependence of these
     background fields. In  gauge theories, there is a unphysical sector
     and consequently unphysical particles may be produced. However, it
     is shown that produced unphysical particles have no contribution to
     backreaction to background fields.

 \end{normalsize}
\end{quote}

\end{titlepage}
\vfil\eject

\section{Introduction}
     Inf\mbox{}lationary  universe scenario is intended to solve some of
     fundamental problems in the standard cosmology such as the horizon,
     f\mbox{}latness and primordial monopole problems \cite{einsa}\cite{guth}
     \cite{sato}. It introduces an exponentially expanding era due to
     the vacuum energy of the inf\mbox{}laton field. Gigantic order of
     expansion can solve two of the above problems ; the horizon and
     primordial monopole problems. It is intuitively expected that
     immediately after inf\mbox{}lation, the vacuum energy is converted
     into radiation energy, and that the Friedmann expansion takes over
     the de Sitter expansion. And at this era, the huge amount of entropy
     is expected to be produced, which solves the f\mbox{}latness problem.

  According to the new inf\mbox{}lationary universe scenario \cite{linde}
  \cite{albst}
  , GUT phase transition
  is of second order. Not only a c-number background metric but also
  a c-number background Higgs field are time-dependent. Relations between
  these c-number background fields and quantum fluctuations
  have been studied for many years.
  In the context of the thermalization after inf\mbox{}lation, several authors
  studied the effective evolution equation of a c-number background
  Higgs field
  \cite{hosa}\cite{morisa}\cite{ringwald}\cite{paz}. However, several important
  problems
  are still left unsolved.

  In this paper, we discuss some aspects of  the consistency of quantum field
  theory when there are spacetime-dependent  c-number background fields.
  In general, particle production can occur in
  spacetime-dependent background. Unphysical particles in gauge theories are
  no exception.
  However, we shall show at the 1-loop level that unphysical particles do
  not contribute
  to backreaction in spite of their condensation. We prove this  fact by using
  the U(1) Higgs model, but extension to more complicated models
  (e.g. SU(5) GUT model) is straightforward.

  This paper is organized as follows: In sect. \ref{sec:lag} we explain our
  U(1) Higgs model. In particular
  we present a gauge fixing condition which is a generalization of the
  familiar $R_{\xi}$ gauge
  to spacetime-dependent background. This plays a crucial role
  in the following
  discussion. In sect. \ref{sec:ext} a free part of the Lagrangian density
  is diagonalized, and mode
  expansion of each component is discussed in sect. \ref{sec:basis}.
  In sect. \ref{sec:vac} we point out that the
  discrepancy between the in-vacuum and the out-vacuum lead to
  the condensation of
  physical and even unphysical particles. However, we can take BRST-invariant
  vacuum states on a certain condition.  In sect. \ref{sec:back} we show that
  the condensation of unphysical particles do not
  contribute to backreaction in the background field equations
  of a physical state. Sect. \ref{sec:sum} is for conclusion.

\setcounter{equation}{0}
\section{Lagrangian\label{sec:lag}}
  We start from the standard Lagrangian density for the U(1) Higgs model
  with a non-minimal
  coupling to the scalar curvature.
  \beq
    \tilag{0}=\df\lag_{0}
  \eeq
  \beq
    \lag_{0}=-\frac14F_{\mu\nu}F^{\mu\nu}+(D_{\mu}\phi)^{\dagger}(D^{\mu}\phi)
    -m^2\phi^{\dagger}
    \phi-\xi R\phi^{\dagger}\phi-\frac{\la}{2}(\phi^{\dagger}\phi)^2
  \eeq
  where
  \beq
    F_{\mu\nu}=\nabla_\mu A_\nu-\nabla_\nu A_\mu=\del_\mu A_\nu-\del_\nu A_\mu
  \eeq
  \beq
    D_\mu\phi=\del_\mu\phi-ieA_\mu\phi
  \eeq
  In this paper we study quantum effects of the gauge and scalar fields
  in the presence of background fields which are spacetime-dependent in
  general. We parametrize the complex scalar field in polar variables and
  shift the modulus field by the classical component \nolinebreak $\phc$,
  \beq
    \phi=\frac1{\sqrt{2}}(\phc+\rho)exp\left[i\frac{\pi}{\phc}\right]
  \eeq

  To avoid the mixing between $A_\mu$ and $\pi$, we adopt the following
  gauge fixing and
  ghost part of the Lagrangian  density:
  \beq
    \tilag{G}=\df\lag_{G}
  \eeq
  \beqa
    \lag_{G}&=&-i\brs\left[-\left(\del^\mu\frac{\cb}{\phc^2}\right)
    \phc^2 A_\mu+\al\cb\left(e\phc\pi
                   +\frac{B}2\right)\right]\\
                 &=&-\left(\del^\mu\frac{B}{\phc^2}\right)\phc^2A_\mu
                 +\al e\phc B\pi+\frac{\al}2B^2\nn\\
                  & &+i\left\{-\phc^2\left(\del_\mu\frac{\cb}{\phc^2}\right)
                  \del^\mu c
                         +\al e^2\phc^2\cb c\right\}
  \eeqa
  where $\brs$ represents the BRST transformation :
  \beq
    \brs A_{\mu}=\del_\mu c
  \eeq
  \beq
    \brs \pi =e\phc c
  \eeq
    \beq
    \brs \cb=iB
  \eeq
  \beq
    \brs(\mbox{otherwise})=0
  \eeq
  (when $g_{\mu\nu}=\eta_{\mu\nu}$ and $\phc=\mbox{const.}$, this gauge
  coincides with
  the $R_\xi$ gauge \cite{fls}\cite{yao}.)

  The total Lagrangian density is hence given by
  \beq
    \tilag{}=\df\;[\lag_{\phc}+\lag_{1}+\lag_{2}+\lag_{I}]
  \eeq
  \beq
    \lag_{\phc}=\frac12(\del_{\mu}\phc)^2-\frac12(m^2+\xi R)\phc^2
    -\frac{\la}8\phc^4
  \eeq
  \beq
    \lag_{1}=\del_{\mu}\phc\del^{\mu}\rho-\left\{(m^2+\xi R)\phc
    +\frac{\la}2\phc^3\right\}\rho
  \eeq
  \beqa
    \lag_{2}&=&\frac12(\del_{\mu}\rho)^2-\frac12\left(m^2+\xi R
    +\frac32\la\phc^2\right)\rho^2\nn\\
                 & &-\frac14F_{\mu\nu}F^{\mu\nu}+\frac12M^2\left(A_\mu
                 -\del_\mu\frac{\pi}{M}
                     \right)^2\nn\\
                & &-M^2\left(\del^\mu\frac{B}{M^2}\right)A_\mu+\al
                 MB\pi+\frac{\al}2B^2\nn\\
                & &+i\left\{-M^2\left(\del_\mu\frac{\cb}{M^2}\right)
                \del^\mu c+\al M^2\cb c\right\}
  \eeqa
  \beqa
    \lag_{I}&=&-\frac{\la}8\rho^4-\frac{\la}2\phc\rho^3\nn\\
               & &+\frac12(e^2\rho^2+2eM\rho)\left(A_\mu-\del_\mu
               \frac{\pi}{M}\right)^2
  \eeqa
  where
  \beq
    M=e\phc
  \eeq
  Note that $M$ is spacetime-dependent in general.

  Field equations  are given by
  \beq
    \Box  A_\mu-\nabla^\nu\nabla_\mu A_\nu+(M+e\rho)^2\left(A_\mu
    -\del_\mu\frac{\pi}{M}
    \right)-M^2\del_\mu\frac{\brB}{M}=0
  \eeq
  \beq
    \frac{\al}{M}(\brB+\pi)+\frac1{M^4}\nabla^\mu (M^2A_\mu)+0
  \eeq
  \beq
    \frac1{M}\nabla^\mu\left\{(M+e\rho)^2\left(A_\mu-\del_\mu
    \frac{\pi}{M}\right)\right\}
    +\al M^2\brB=0
  \eeq
  \beq
    \frac1{M}\nabla^\mu\left\{M^2\left(\del_\mu\frac{\brc}{M}\right)
    \right\}+\al M^2\brc=0
  \eeq
  \beq
    \frac1{M}\nabla^\mu\left\{M^2\left(\del_\mu\frac{\brcb}{M}\right)
    \right\}+\al M^2\brcb=0
  \eeq
  where
  \beq
    \brB=\frac{B}{M}\quad,\qquad\brc=Mc\quad,\qquad\brcb=\frac{\cb}{M}
  \eeq
  When the classical component $\phc$ is  a solution of the following
  classical field  equation:
  \beq
    \Box\phc+(m^2+\xi R)\phc+\frac{\la}2\phc^3=0
  \eeq
  $ \lag_{1}$ vanishes. The  field equation of $\rho$ is then given by
  \beq
    \Box\rho+\left(m^2+\xi R+\frac32\la\phc^2\right)\rho+\frac{\la}2\rho^3
    +\frac32\la\phc\rho^2
    -e(M+e\rho)\left(A_\mu-\del_{\mu}\frac{\pi}{M}\right)^2=0
  \eeq

\pagebreak
  $\tilag{}$ is invariant under the BRST transformation. The conserved
  Neother current is
  \beq
    j_{B\:\mu}=\frac1{\df}\left\{\frac{\del\tilag{}}{\del\,\nabla^\mu
     A^\nu}\del^\nu c+\frac{\del\tilag{}}
     {\del\,\del^\mu\pi}Mc+iB(\del/\del\,\del^\mu\cb)\tilag{}\right\}
  \eeq
  from which the conserved charge is given as
  \beqa
    \bolQB&=&\ind{3}{x}\:\df \;j_{B}^{\:0}\nn\\
               &=&\ind{3}{x}\:\df\;\{\brB\:\ddel^{0}\brc\}
  \eeqa
  with the help of the field equations. $\tilag{}$ is also invariant under
  the scale transformation:
  \beq
    c\quad\to\quad e^\th c
  \eeq
  \beq
    \cb\quad\to\quad e^{-\th}\cb
  \eeq
  The conserved current and the charge of this transformation are given by
  \beqa
    j_{c\:\mu}&=&\frac1{\df}\left\{(\del\tilag{}/\del\,\del^\mu c)c
    +(-\cb)(\del/\del\,\del^\mu\cb)\tilag{}
                          \right\}\nn\\
                   &=&i\{\brcb\:\ddel_\mu \brc\}
  \eeqa

  \beqa
    \bolQc&=&\ind{3}{x}\:\df\;j_{c}^{\:0}\nn\\
               &=&i\ind{3}{x}\:\df\;\{\brcb\:\ddel^{0}\brc\}
  \eeqa
  where
  \beq
    \varphi\:\ddel_{\mu}\psi=\varphi\:\del_{\mu}\psi-(\del_{\mu}\varphi)\,\psi
  \eeq

  Canonical momenta $\Pi_{\rho},\;\Pi^i,\;\Pi_{B},\;\Pi_{\pi},\;\Pi_{c}$ and
  $\Pi_{\cb}$ conjugate to $\rho,\; A_{i},\; B,\;\pi,\; c$ and $\cb$,
   respectively,
  are
  \beq
    \Pi_{\rho}=\df\;\del^{0}\rho
  \eeq
  \beq
    \Pi^{i}=\df\;(\del^{i}A^{0}-\del^{0}A^{i})
  \eeq
  \beq
    \Pi_{B}=-\df\;A^{0}
  \eeq
  \beq

\Pi_{\pi}=\frac{\df}{M}\;(M+e\rho)^2\left(\del^{0}\frac{\pi}{M}-A^{0}\right)
  \eeq
  \beq
    \Pi_{c}=-i\df\;M^2\del^{0}\frac{\cb}{M^2}
  \eeq
  \beq
    \Pi_{\cb}=i\df\;\del^{0}c
  \eeq
  We set up Canonical (anti-)commutation relations as
  \beq
    [\Pi_{\rho}(x)\;,\;\rho(y)]_{\mbox{E.T.}}=-i\de^3(x-y)
  \eeq
  \beq
    [\Pi^{i}(x)\;,\;A_{j}(y)]_{\mbox{E.T.}}=-i\de^i_{\;j}\de^3(x-y)
  \eeq
  \beq
    [\Pi_{B}(x)\;,\;B(y)]_{\mbox{E.T.}}=-i\de^3(x-y)
  \eeq
  \beq
    [\Pi_{\pi}(x)\;,\;\pi(y)]_{\mbox{E.T.}}=-i\de^3(x-y)
  \eeq
  \beq
    \{\Pi_{c}(x)\;,\;c(y)\}_{\mbox{E.T.}}=-i\de^3(x-y)
  \eeq
  \beq
    \{\Pi_{\cb}(x)\;,\;\cb (y)\}_{\mbox{E.T.}}=-i\de^3(x-y)
  \eeq

\setcounter{equation}{0}
\section{Extraction of a massive vector field\label{sec:ext}}
  The Laglangian density $\tilag{}$ should be diagonalized. A physical
  massive vector field,
  which is denoted as $U_{\mu}$, is a linear combination of $A_{\mu},\,
  \pi$ and $B$.
  \beq
    U_{\mu}=A_{\mu}-\del_{\mu}\frac{\pi}{M}-\del_{\mu}\frac{\brB}{M}
  \eeq
  From the BRST transformation of $A_{\mu},\,\pi$ and $B$, $\:U_{\mu}$
  is shown to be BRST-invariant. In terms of $U_{\mu}$,
  the Laglangian density $\tilag{}$ can be rewritten as
  \beq
    \tilag{}=\df\;[\lag_{\phc}+\lag_{1}+\lag_{2}^{\prime}+\lag_{I}^{\prime}]
  \eeq

\pagebreak
  \beqa
    \lag_{2}^{\prime}&=&\frac12(\del_{\mu}\rho)^2-\frac12\left(m^2
    +\xi R+\frac32\la\phc^2
                                     \right)\rho^2\nn\\
                              & &-\frac14H_{\mu\nu}H^{\mu\nu}
                              +\frac12M^2U_{\mu}U^{\mu}\nn\\
                              & &+\frac12\left\{-M^2\left(\del_{\mu}
                              \frac{\brB}{M}\right)\left(\del^{\mu}
                              \frac{\brB}{M}\right)+\al M^2\brB^2\right\}\nn\\
                              & &+\left\{-M^2\left(\del_{\mu}
                              \frac{\brB}{M}\right)\left(\del^{\mu}
                              \frac{\pi}{M}\right)+\al M^2\brB\pi\right\}\nn\\
                              & &+i \left\{-M^2\left(\del_{\mu}
                              \frac{\brcb}{M}\right)\left(\del^{\mu}
                              \frac{\brc}{M}\right)+\al M^2\brcb\brc\right\}
  \eeqa
  \beq
    \lag_{I}^{\prime}=-\frac{\la}8\rho^4-\frac{\la}2\phc\rho^3
    +\frac12(e^2\rho^2+2eM\rho)
                               \left(U_{\mu}+\del_{\mu}\frac{\brB}{M}\right)^2
  \eeq
  where
  \beq
    H_{\mu\nu}=\nabla_{\mu}U_{\nu}-\nabla_{\nu}U_{\mu}
    =\del_{\mu}U_{\nu}-\del_{\nu}U_{\mu}=F_{\mu\nu}
  \eeq

  In the rest of this paper we shall be mainly concerned with effects
  of spacetime dependence
  in the background scalar field $\phc$ and the background metric
  $g_{\mu\nu}$ to these free fields.
  The interaction part of  the Lagrangian density can be handled as
  a perturbation but shall not
  be studied in detail. Free field equations derived from the bilinear
  part of the Laglangian density
  are hence given by
  \beq
    \Box U_{\mu}-\nabla^{\nu}\nabla_{\mu}U_{\nu}+M^2U_{\mu}=0
  \eeq
  \beq
    \Box\brB+\left(\al M^2-\frac{\Box M}{M}\right) \brB=0
  \eeq
  \beq
    \Box\pi+\left(\al M^2-\frac{\Box M}{M}\right) \pi=0
  \eeq
  \beq
    \Box\brc+\left(\al M^2-\frac{\Box M}{M}\right) \brc=0
  \eeq
  \beq
    \Box\brcb+\left(\al M^2-\frac{\Box M}{M}\right) \brcb=0
  \eeq

  When $\phc$ is a solution of the classical field equation, the field equation
  of $\rho$ becomes
   \beq
    \Box\rho+\left(m^2+\xi R+\frac32\la\phc^2\right) \rho=0
  \eeq

  At this level, each free field corresponds to a one particle state.
  Therefore we call $\rho$ and
  $U_{\mu}\;$  physical fields in strong sense and call $B,\,\pi,\,c$
  and $\cb\;$  a BRST quartet.

\setcounter{equation}{0}
\section{Basis functions and innerproducts\label{sec:basis}}
  We expand the quantum fields in normal modes.
  \beq
    U_{\mu}(x)=\sum_{k,\: a}\left\{\bolU(ka)\:f_{\mu}(ka|x)
    +\bolUdag(ka)\:f^{\ast}_{\mu}(ka|x)
                        \right\}
  \eeq
  \beq
    B(x)=M(x)\sum_{k}\left\{\bolB(k)\:g_{B}(k|x)
    +\bolBdag(k)\:g^{\ast}_{B}(k|x)
                        \right\}
  \eeq
  \beq
    \pi(x)=\sum_{k}\left\{\bolpi(k)\:g_{\pi}(k|x)
    +\bolpidag(k)\:g^{\ast}_{\pi}(k|x)
                        \right\}
  \eeq
  \beq
    c(x)=\frac1{M(x)}\sum_{k}\left\{\bolc(k)\:g_{c}(k|x)
    +\bolcdag(k)\:g^{\ast}_{c}(k|x)
                        \right\}
  \eeq
  \beq
    \cb(x)=M(x)\sum_{k}\left\{\bolcb(k)\:g_{\cb}(k|x)
    +\bolcbdag(k)\:g^{\ast}_{\cb}(k|x)
                        \right\}
  \eeq
  \beq
    \rho(x)=\sum_{k}\left\{\bolrh(k)\:h(k|x)+\bolrhdag(k)\:h^{\ast}(k|x)
                        \right\}
  \eeq
  Here $k$ and $a$ are a set of quantum numbers to label modes.
  In Minkowski space, they are related to momentum and spin . ($a=1,\;2,\;3$)

  Basis functions satisfy the following wave equations:
  \beq
    \left\{
      \begin{array}{ll}
        \Box f_{\mu}-\nabla^{\nu}\nabla_{\mu}f_{\nu}+M^2f_{\mu}=0\\
        \nabla^{\mu}(M^2f_{\mu})=0
      \end{array}
    \right.
  \eeq
  \beq
    \Box \gF+\left(\al M^2-\frac{\Box M}{M}\right)\gF=0
  \eeq
  \beq
    \Box h+\left(m^2+\xi R+\frac32\la\phc\right)h=0
  \eeq
  where $F=B,\;\pi,\;c,\;\cb$

  Let us introduce innerproducts for basis functions. We can show that
  \beqa
    \nabla_{\nu}[f_{\mu}^{\ast}(ka|x)\{\nabla^{\nu}f^{\mu}(\kp\ap|x)
    &- &\nabla^{\mu}f^{\nu}
    (\kp\ap|x)\}\nn\\
      &- &\{\nabla^{\nu}f^{\mu\ast}(ka|x)-\nabla^{\mu}f^{\nu\ast}(ka|x)\}
    f_{\mu}(\kp\ap|x)]=0
  \eeqa
  \beq
    \nabla_{\nu}[g_{B}^{\ast}(k|x)\ddel^{\nu}g_{\pi}(\kp|x)]=0
  \eeq
  \beq
    \nabla_{\nu}[g_{\cb}^{\ast}(k|x)\ddel^{\nu}g_{c}(\kp|x)]=0
  \eeq
  \beq
    \nabla_{\nu}[h^{\ast}(k|x)\ddel^{\nu}h(\kp|x)]=0
  \eeq
  by using the wave equations. Thus we can define the time-independent
  innerproducts as
  \beqa
    \dlangle f(ka),\:f(\kp\ap)\drangle&=&-i\ind{3}{x}\:\df\;
    [f_{\mu}^{\ast}(ka|x)\{\nabla^{0}
      f^{\mu}(\kp\ap|x) -\nabla^{\mu}f^{0}(\kp\ap|x)\}\nn\\
     & &\mbox{ } -\{\nabla^{0}f^{\mu\ast}(ka|x)
     -\nabla^{\mu}f^{0\ast}(ka|x)\}f_{\mu}(\kp\ap|x)]
  \eeqa
  \beq
    \dlangle g_{B}(k),\:g_{\pi}(\kp)\drangle=i\ind{3}{x}\:\df\;[g_{B}^{\ast}
    (k|x)\ddel^{0}g_{\pi}(\kp|x)]
  \eeq
  \beq
    \dlangle g_{\cb}(k),\:g_{c}(\kp)\drangle=i\ind{3}{x}\:\df\;[g_{\cb}^{\ast}
    (k|x)\ddel^{0}g_{c}(\kp|x)]
  \eeq
  \beq
    \dlangle h(k),\:h(\kp)\drangle=i\ind{3}{x}\:\df\;[h^{\ast}(k|x)
    \ddel^{0}h(\kp|x)]
  \eeq
  We impose the following orthonormality:
  \beq
    \dlangle f(ka),\:f(\kp\ap)\drangle=\de(k,\:\kp)\de(a,\:\ap),\quad
    \dlangle  f^{\ast}(ka),\:f(\kp\ap)\drangle=0
  \eeq
  \beq
    \dlangle g_{B}(k),\:g_{\pi}(\kp)\drangle=\de(k,\:\kp),\quad
    \dlangle g_{B}^{\ast}(k),\:g_{\pi}(\kp)\drangle=0
  \eeq
  \beq
    \dlangle g_{\cb}(k),\:g_{c}(\kp)\drangle=\de(k,\:\kp),\quad
    \dlangle g_{\cb}^{\ast}(k),\:g_{c}(\kp)\drangle=0
  \eeq
  \beq
    \dlangle h(k),\:h(\kp)\drangle=\de(k,\:\kp),\quad
    \dlangle h^{\ast}(k),\:h(\kp)\drangle=0
  \eeq

  By using the innerproducts, we can express the coefficients as
  \beq
    \bolU(ka)=\dlangle f(ka),\:U\drangle ,\quad \bolUdag(ka)
    =-\dlangle f^{\ast}(ka),\:U\drangle
  \eeq
  \beq
    \bolB(k)=\dlangle g_{\pi}(k),\:B\drangle ,
    \quad \bolBdag(k)=-\dlangle g_{\pi}^{\ast}(k),\:B\drangle
  \eeq
  \beq
    \bolpi(k)=\dlangle g_{B}(k),\:\pi\drangle ,\quad
    \bolpidag(k)=-\dlangle g_{B}^{\ast}(k),\:\pi\drangle
  \eeq
  \beq
    \bolc(k)=\dlangle g_{\cb}(k),\:c\drangle ,
    \quad \bolcdag(k)=-\dlangle g_{\cb}^{\ast}(k),\:c\drangle
  \eeq
  \beq
    \bolcb(k)=\dlangle g_{c}(k),\:\cb\drangle ,\quad
    \bolcbdag(k)=-\dlangle g_{c}^{\ast}(k),\:\cb\drangle
  \eeq
  \beq
    \bolrh(k)=\dlangle h(k),\:\rho\drangle ,\quad
     \bolrhdag(k)=-\dlangle h^{\ast}(k),\:\rho\drangle
  \eeq
  From the canonical (anti-)commutation relations, we obtain
  (anti-)commutation relations
   among coefficients.
  \beq
    [\bolU(ka),\:\bolUdag(\kp\ap)]=\de(k,\:\kp)\de(a,\:\ap)
  \eeq
  \beq
    [\bolB(k),\:\bolpidag(\kp)]=[\bolpi(k),\:\bolBdag]=-\de(k,\:\kp)
  \eeq
  \beq
    [\bolpi(k),\:\bolpidag(\kp)]=\de(k,\:\kp)
  \eeq
  \beq
    \{\bolc(k),\:\bolcbdag(\kp)\}=-\{\bolcdag(k),\:\bolcb(\kp)\}=i\de(k,\:\kp)
  \eeq
  \beq
    [\bolrh(k),\:\bolrhdag(\kp)]=\de(k,\:\kp)
  \eeq
  \beq
    \mbox{otherwise}=0
  \eeq

\setcounter{equation}{0}
\section{Vacuum states\label{sec:vac}}
  In this section, we shall construct the "in Fock space" $\Vi$ and
  the "out Fock space" $\Vo$, and investigate the relation between them.

  If basis functions $f, \:\gF$ and $h$ are positive frequency solutions
  in the region $t \rightarrow -\infty$, we write them as $f_{in},\:\gF_{in}$
  and $h_{in}$. In a parallel way, we can define another basis, $f_{out}
  ,\:\gF_{out}$ and $h_{out}$ which are positive frequency solutions
  in the region $t \rightarrow \infty$. When there is a unique positive
  frequency solution, we have the following relation:
  \beq
    g_{in}\equiv g_{B\,in}=g_{\pi\,in}=g_{c\,in}=g_{\cb\,in}
     \label{eqn:gin}
  \eeq
 because they satisfy the same equation and the same boundary condition.
 Similarly, we also have
  \beq
    g_{out}\equiv g_{B\,out}=g_{\pi\,out}=g_{c\,out}=g_{\cb\,out}
  \eeq
  But, when the positive frequency solutions are not unique, it is not obvious
  what condition should be imposed. We shall pick out this from
  the BRST-invariance of vacuum states.

  As is well known, the basis $\{f_{in},\:\gF_{in},\:h_{in}\:;\:f^{\ast}_{in},
  \:\gF^{\ast}_{in}
  ,\:h^{\ast}_{in}\}$ are in general different from the basis
   $\{f_{out},\:\gF_{out},\:h_{out}\:;\:f^{\ast}_{out},\:\gF^{\ast}_{out},
   \:h^{\ast}_{out}\}$
   in  the presence of spacetime-dependent background fields.
  We can expand the fields in two ways. Using the orthonormality of basis
  functions,  the
  coefficients of $\{f_{in},\:\gF_{in},\:h_{in}\:;\:f^{\ast}_{in},
  \:\gF^{\ast}_{in},\:h^{\ast}_{in}\}$
  are related with the coefficients of  $\{f_{out},\:\gF_{out},\:h_{out}\:;
  \:f^{\ast}_{out},\:\gF^{\ast}_{out},\:h^{\ast}_{out}\}$ by the following
  Bogoliubov transformation:
  \beq
    \bolU_{in}(ka)=\sum_{\kp,\:\ap}\left\{\alU^{\ast}(ka,\kp\ap)
                           \: \bolU_{out}(\kp\ap)
                            -\beU^{\ast}(ka,\kp\ap)\:\bolUdag_{out}(\kp\ap)
                            \right\}
  \eeq
  \beq
    \bolB_{in}(k)=\sum_{\kp}\left\{ \balBp^{\ast}(k,\kp)
                           \:\bolB_{out}(\kp)
                            -\bbeBp^{\ast}(k,\kp)\:\bolBdag_{out}(\kp)\right\}
  \eeq
  \beq
    \bolpi_{in}(k)=\sum_{\kp}\left\{ \alBp^{\ast}(k,\kp)
                           \:\bolpi_{out}(\kp)
                            -\beBp^{\ast}(k,\kp)\:\bolpidag_{out}(\kp)\right\}
  \eeq
  \beq
    \bolc_{in}(k)=\sum_{\kp}\left\{\alcc^{\ast}(k,\kp) \:\bolc_{out}(\kp)
                            -\becc^{\ast}(k,\kp)\:\bolcdag_{out}(\kp)\right\}
  \eeq
  \beq
    \bolcb_{in}(k)=\sum_{\kp}\left\{\balcc^{\ast}(k,\kp)
                           \:\bolcb_{out}(\kp)
                            -\bbecc^{\ast}(k,\kp)\:\bolcbdag_{out}(\kp)\right\}
  \eeq
  \beq
    \bolrh_{in}(k)=\sum_{\kp}\left\{\alr^{\ast}(k,\kp)
                           \:\bolrh_{out}(\kp)
                            -\ber^{\ast}(k,\kp)\:\bolrhdag_{out}(\kp)\right\}
  \eeq
  where
  \beq
    \alU(ka,\kp\ap)=\dlangle f_{out}(\kp\ap),\:f_{in}(ka)\drangle
  \eeq
  \beq
    \beU(ka,\kp\ap)=-\dlangle f^{\ast}_{out}(\kp\ap),\:f_{in}(ka)\drangle
  \eeq
  \beq
    \alBp(k,\kp)=\dlangle g_{\pi\:out}(\kp),\:g_{B\:in}(k)\drangle
  \eeq
  \beq
    \beBp(k,\kp)=-\dlangle g^{\ast}_{\pi\:out}(\kp),\:g_{B\:in}(k)\drangle
  \eeq
  \beq
    \balBp(k,\kp)=\dlangle g_{B\:out}(\kp),\:g_{\pi\:in}(k)\drangle
  \eeq
  \beq
    \bbeBp(k,\kp)=-\dlangle g^{\ast}_{B\:out}(\kp),\:g_{\pi\:in}(k)\drangle
  \eeq
  \beq
    \alcc(k,\kp)=\dlangle g_{c\:out}(\kp),\:g_{\cb\:in}(k)\drangle
  \eeq
  \beq
    \becc(k,\kp)=-\dlangle g^{\ast}_{c\:out}(\kp),\:g_{\cb\:in}(k)\drangle
  \eeq
  \beq
    \balcc(k,\kp)=\dlangle g_{\cb\:out}(\kp),\:g_{c\:in}(k)\drangle
  \eeq
  \beq
    \bbecc(k,\kp)=-\dlangle g^{\ast}_{\cb\:out}(\kp),\:g_{c\:in}(k)\drangle
  \eeq
  \beq
    \alr(k,\kp)=\dlangle h_{out}(\kp),\:h_{in}(k)\drangle
  \eeq
  \beq
    \ber(k,\kp)=-\dlangle h^{\ast}_{out}(\kp),\:h_{in}(k)\drangle
  \eeq

  Now we can construct the  "in Fock space" $\Vi$ and  the "out Fock space"
  $\Vo$. The in-vacuum is characterized by
  \beq
    \bolU_{in}|0_{in}\rangle=\bolB_{in}|0_{in}\rangle=\bolpi_{in}|0_{in}
    \rangle=\bolc_{in}|0_{in}\rangle=\bolcb_{in}|0_{in}\rangle
    =\bolrh_{in}|0_{in}\rangle=0
  \eeq
  The in Fock space is obtained by applying in-creation operators to
  $|0_{in}\rangle$.
  \beq
    \Vi=\left\{\cdots\bolUdag_{in}\cdots\bolBdag_{in}\cdots\bolpidag_{in}
    \cdots\bolcdag_{in}
    \cdots\bolcbdag_{in}\cdots\bolrhdag_{in}\cdots|0_{in}\rangle\right\}
  \eeq
  Similarly the out-vacuum is defined by
  \beq
    \bolU_{out}|0_{out}\rangle=\bolB_{out}|0_{out}\rangle
    =\bolpi_{out}|0_{out}\rangle
    =\bolc_{out}|0_{out}\rangle=\bolcb_{out}|0_{out}\rangle
    =\bolrh_{out}|0_{out}\rangle=0
  \eeq
  and the out Fock space is
  \beq
    \Vo=\left\{\cdots\bolUdag_{out}\cdots\bolBdag_{out}\cdots\bolpidag_{out}
    \cdots\bolcdag_{out}
    \cdots\bolcbdag_{out}\cdots\bolrhdag_{out}\cdots|0_{out}\rangle\right\}
  \eeq

  In general, $|0_{in}\rangle\neq |0_{out}\rangle$. Because the coefficients
  of  $\{f_{in},\:\gF_{in},\:h_{in}\:;\:f^{\ast}_{in},\:\gF^{\ast}_{in},
  \:h^{\ast}_{in}\}$ are related with the coefficients of $\{f_{out},
  \:\gF_{out},\:h_{out}\:;\:f^{\ast}_{out},\:\gF^{\ast}_{out},
  \:h^{\ast}_{out}\}$ by the Bogoliubov transformation,
  the in-vacuum can be expressed in terms of  the
  out-vacuum.

  \pagebreak
  From the definition of the in-vacuum,
  \beqa
    |0_{in}\rangle\propto& &exp\left[\sum_{\{k,\,a\}\:\{\kp,\,\ap\}}\frac12
    \la^{\ast}_{U}(ka,\kp\ap)
    \bolUdag_{out}(ka)\bolUdag_{out}(\kp\ap)\right.\nn\\
    & &-\sum_{k,\:\kp}\la^{\ast}_{B\,\pi}(k,\:\kp)\left\{\frac12
    \bolBdag_{out}(k)\bolBdag_{out}(\kp)+\bolBdag_{out}(k)
    \bolpidag_{out}(\kp)\right\}\nn\\
    & &-\sum_{k,\:\kp}i\la^{\ast}_{\cb\,c}(k,\:\kp)\bolcbdag_{out}(k)
    \bolcdag_{out}(\kp)\nn\\
    & &+\left.\sum_{k,\:\kp}\frac12\la^{\ast}_{\rho}(k,\:\kp)\bolrhdag_{out}(k)
    \bolrhdag_{out}(\kp) \right]|0_{out}\rangle
  \label{eqn:invacuum}
  \eeqa
  where
  \beq
    \la_{U}(ka,\kp\ap)=(\alU^{-1}\beU)(ka,\kp\ap)
  \eeq
  \beq
    \la_{B\,\pi}(k,\kp)=(\alBp^{-1}\beBp)(k,\kp)
    =(\balBp^{-1}\bbeBp)(\kp,k)=\bar{\la}_{B\,\pi}(\kp,k)
  \eeq
  \beq
    \la_{\cb\,c}(k,\kp)=(\alcc^{-1}\becc)(k,\kp)
    =(\balcc^{-1}\bbecc)(\kp,k)=\bar{\la}_{\cb\,c}(\kp,k)
  \eeq
  \beq
    \la_{\rho}(k,\kp)=(\alr^{-1}\ber)(k,\kp)
  \eeq
  (From the unitarity condition of  the Bogoliubov coefficients,
   $\la_{U},\:\la_{B\,\pi},\:\bar{\la}_{B\,\pi}$ and $\la_{\rho}$ are
  symmetric. i.e., $\la(k,\kp)=\la(\kp,k)$ .
  The proof is given in Appendix.)
  This formula is the generalization of that given by \cite{wada} in which RW
  metric is assumed
  and the (anti-)commutation relations of field operators are diagonal.

  From this formula, the in-vacuum can be regarded as the state in which
  out-particles including unphysical particles condense on the out-vacuum.
  One might worry about
  that the in-vacuum could be a unphysical state in the system described by
  the bilinear part of the Laglangian density. In the rest of this section,
  we shall show what condition should be chosen for BRST-invariant vacuum
  states.

  Now we follow the formalism given in \cite{kugoji}, in which the physical
  states are selected by
  the subsidiary condition:
  \beq
    \bolQB|phys.\rangle=0
  \eeq
  The BRST charge in terms of annihilation and creation operators is
  \beqa
    \bolQB_{in}=(-i)&\displaystyle{\sum_{k,\,\kp}}&\left\{
                  \bolB_{in(out)}(k)
                  \dlangle g^{\ast}_{B\:in(out)}(k),
                  \:g^{\ast}_{c\:in(out)}(\kp)\drangle
                  \bolcdag_{in(out)}(\kp)\right.\nn\\
                &+&\bolBdag_{in(out)}(k)
                  \dlangle g_{B\:in(out)}(k),
                  \:g_{c\:in(out)}(\kp)\drangle
                  \bolc_{in(out)}(\kp)\nn\\
                &+&\bolB_{in(out)}(k)
                  \dlangle g^{\ast}_{B\:in(out)}(k),
                  \:g_{c\:in(out)}(\kp)\drangle
                  \bolc_{in(out)}(\kp)\nn\\
                &+&\left.\bolBdag_{in(out)}(k)
                  \dlangle g_{B\:in(out)}(k),
                  \:g^{\ast}_{c\:in(out)}(\kp)\drangle
                  \bolcdag_{in(out)}(\kp)
                 \right\}
  \eeqa
  So that the in(out)-vacuum is a physical state, the following condition is
  necessary and sufficient.
  \beqa
    \exists z_{in(out)}(k,\kp)&:&\;det\:z_{in(out)}\ne 0 \nn\\
    & &\left\{
    \begin{array}{ll}
      g_{B\:in(out)}(k)=\displaystyle{\sum_{\kp}}\:z^{\ast}_{in(out)}(k,\kp)
                      g_{\cb\:in(out)}(\kp)\\
      g_{\pi\:in(out)}(k)=\displaystyle{\sum_{\kp}}\:z^{-1}_{in(out)}(\kp,k)
                      g_{c\:in(out)}(\kp)
    \end{array}
    \right.
  \label{eqn:condition}
  \eeqa
  If both vacuum states are physical, we have the following relation:
  \beq
    \bolQB=\bolQB_{in}=\bolQB_{out}
  \eeq
  where
  \beq
    \bolQB_{in}= i\sum_{k,\,\kp}\left\{\bolB_{in}(k)
                  z^{\ast}_{in}(k,\kp)\bolcdag_{in}(\kp)
                 -\bolBdag_{in}(k)
                  z_{in}(k,\kp)\bolc_{in}(\kp)
                 \right\}
  \eeq
  \beq
    \bolQB_{out}= i\sum_{k,\,\kp}\left\{\bolB_{out}(k)
                  z^{\ast}_{out}(k,\kp)\bolcdag_{out}(\kp)
                 -\bolBdag_{out}(k)
                  z_{out}(k,\kp)\bolc_{out}(\kp)
                 \right\}
  \eeq

  In case $z_{in}=z_{out}=I$, the above formula was proved in \cite{hoho}
  using the unitarity condition of the Bogoliubov coefficients.
  The authors of \cite{hoho} concluded that the invariance of the BRST charge
  under the Bogoliubov transformation implies an absence of
  unphysical particles.
  However, we would rather claim that unphysical particles are condensed but
  do not contribute to physical processes such as backreaction to background
  fields, which we will prove in the next section.

  Before completing this section, we shall again look the form of condensation
  when both vacuum states are physical. From (\ref{eqn:invacuum}) and
  (\ref{eqn:condition}),
  \beqa
    |0_{in}\rangle&\propto&exp\left[\sum_{\{k,\,a\}\:\{\kp,\,\ap\}}\frac12
    \la^{\ast}_{U}(ka,\kp\ap)\bolUdag_{out}(ka)\bolUdag_{out}(\kp\ap)\right.
    \nn\\
    & &+\sum_{k,\:\kp}\la^{\ast}_{BRST}(k,\:\kp)\left\{i\bolQB_{out},
    \:i\bolcbdag_{out}(k)
    \left(\bolpidag_{out}(\kp)+\frac12\bolBdag_{out}(\kp)\right)
    \right\}\nn\\
    & &+\left.\sum_{k,\:\kp}\frac12\la^{\ast}_{\rho}(k,\:\kp)\bolrhdag_{out}(k)
    \bolrhdag_{out}(\kp) \right]|0_{out}\rangle
  \eeqa
  where
  \beq
    \la_{BRST}(k,\kp)=\sum_{\kpp}z_{out}^{\ast\,-1}(k,\kpp)\la_{B\pi}(\kpp,\kp)
                     =\sum_{\kpp}\la_{\cb c}(k,\kpp)z_{out}^{-1}(\kpp,\kp)
  \eeq

  Consequently, in case both vacuum states are BRST-invariant, the condensation
  of the BRST quartet sector occur in a BRST-exact form and we must choose
basis functions as
  \beq
    \left\{
    \begin{array}{ll}
      g_{B\:in}=z^{\ast}_{in}g_{\cb\:in}
      ,\quad z^{T}_{in}g_{\pi\:in}=g_{c\:in}\\
      g_{B\:out}=z^{\ast}_{in}g_{\cb\:out}
      ,\quad z^{T}_{out}g_{\pi\:out}=g_{c\:out}
    \end{array}
    \right.
  \eeq
  where we use matrix representation.

\setcounter{equation}{0}
\section{Backreaction\label{sec:back}}
  Now we shall prove the fact that produced unphysical particles do not
  contribute to backreaction. We look at backreaction in the Einstein equations
  first. And next, backreaction in the effective equation of the classical
  component $\phc$ will be considered.

  \subsection{The Einstein equations}
    The energy-momentum tensor  derived from the bilinear part of
     the Laglangian density is given by
    \beq
      \EM{2}=\EM{U_{\mu}}+\EM{\rho}+\EM{B\pi}+\EM{\cb c}
    \eeq
    where
    \beq
      \EM{U_{\mu}}=-H_{\mu\si}H_{\nu}^{\;\:\si}+M^2U_{\mu}U_{\nu}
                   +\frac14g_{\mu\nu}\left(H_{\ga\si}H^{\ga\si}
                   -2M^2U_{\si}U^{\si}\right)
    \eeq
    \beqa
       \EM{\rho}&=&\del_{\mu}\rho\del_{\nu}\rho+\frac12g_{\mu\nu}
                   \left\{(m^2+\xi R+\frac32\la\phc^2)\rho^2
                   -\del_{\si}\rho\del^{\si}\rho\right\}\nn\\
                & &+2\xi\left\{g_{\mu\nu}(\rho\Box\rho
                   +\del_{\si}\rho\del^{\si}\rho)
                   -\frac12\rho(\nabla_{\mu}\del_{\nu}
                   +\nabla_{\nu}\del_{\mu})\rho
                   -\del_{\mu}\rho\del_{\nu}\rho
                   -\frac12R_{\mu\nu}\rho^2\right\}
    \eeqa
    \beqa
      \EM{B\pi}&=&-M^2\left\{\del_{\mu}\frac{\brB}{M}\del_{\nu}
                  \frac{\brB}{M}+\frac12g_{\mu\nu}\left(\al\brB^2
                  -\del_{\si}\frac{\brB}{M}\del^{\si}\frac{\brB}{M}
                  \right)\right\}\nn\\
               & &-M^2\left\{\del_{\mu}\frac{\brB}{M}\del_{\nu}\frac{\pi}{M}
                  +\del_{\nu}\frac{\brB}{M}\del_{\mu}\frac{\pi}{M}
                  +g_{\mu\nu}\left(\al\brB\pi-\del_{\si}\frac{\brB}{M}
                  \del^{\si}\frac{\pi}{M}\right)\right\}
    \eeqa
    \beq
       \EM{\cb c}=-iM^2\left\{\del_{\mu}\frac{\brcb}{M}\del_{\nu}
                  \frac{\brc}{M}+\del_{\nu}\frac{\brcb}{M}\del_{\mu}
                  \frac{\brc}{M}+g_{\mu\nu}\left(\al\brcb\brc
                  -\del_{\si}\frac{\brcb}{M}\del^{\si}\frac{\brc}{M}
                  \right)\right\}
    \eeq

    The Einstein equations including 1-loop backreaction are
    \beq
      G_{\mu\nu}=-\ka\left\{\EM{\phc}+\langle\EM{2}\rangle\right\}
    \eeq
    where
    \beq
      G_{\mu\nu}=R_{\mu\nu}-\frac12g_{\mu\nu}R
    \eeq
    \beqa
       \EM{\phc}&=&\del_{\mu}\phc\del_{\nu}\phc+\frac12g_{\mu\nu}
                   \left\{(m^2+\xi R)\phc^2+\frac{\la}4\phc^4
                   -\del_{\si}\phc\del^{\si}\phc\right\}\nn\\
                & &+2\xi\left\{g_{\mu\nu}(\phc\Box\phc+\del_{\si}
                   \phc\del^{\si}\phc)-\frac12\phc(\nabla_{\mu}\del_{\nu}
                   +\nabla_{\nu}\del_{\mu})\phc
                   \right.\nn\\
                & &\qquad\qquad\qquad\qquad\qquad\qquad\qquad
                   \left.-\del_{\mu}\phc\del_{\nu}\phc
                   -\frac12R_{\mu\nu}\phc^2\right\}
    \eeqa

    In general, the contributions from $\langle\EM{B\pi}\rangle$ and
    $\langle\EM{\cb c}\rangle$ don't vanish. But if the state is physical,
    they cancel each other because $\EM{B\pi}+\EM{\cb c}$ is a BRST-exact
    operator.
    \beq
      \EM{BRST}\equiv\EM{B\pi}+\EM{\cb c}=\left\{i\bolQB,
      \:\Upsilon_{BRST\:\mu\nu}\right\}
    \eeq
    where
    \beqa
      \Upsilon_{BRST\:\mu\nu}&=&iM^2\left[\del_{\mu}\frac{\brcb}{M}
                                \del_{\nu}\left(\frac{\pi}{M}
                                +\frac12\frac{\brB}{M}\right)
                                +\del_{\nu}\frac{\brcb}{M}\del_{\mu}
                                \left(\frac{\pi}{M}
                                +\frac12\frac{\brB}{M}\right)\right.\nn\\
                             & &\left.+g_{\mu\nu}\left\{\al\brcb
                                \left(\pi+\frac12\brB\right)
                                -\del_{\si}\frac{\brcb}{M}\del^{\si}
                                \left(\frac{\pi}{M}
                                +\frac12\frac{\brB}{M}\right)\right\}\right]
    \eeqa
    \nopagebreak
    Consequently, there is no backreaction from the produced BRST quartet.

    \pagebreak
    Especially  when the in-vacuum is physical, the cancellation can be
    explicitly shown as follows,
    \beqa
      \langle0_{in}|\EM{BRST}|0_{in}\rangle
      &=&\sum_{k}M^2\left[\left(\del_{\mu}\frac{g_{B\:in}(k)}{M}
       \right)\left(\del_{\nu}\frac{g^{\ast}_{\pi\:in}(k)}{M}\right)
       +\left(\del_{\nu}\frac{g_{B\:in}(k)}{M}\right)
       \left(\del_{\mu}\frac{g^{\ast}_{\pi\:in}(k)}{M}\right)\right.\nn\\
      & &\qquad\quad+\left.g_{\mu\nu}\left\{\al \:g_{B\:in}(k)
       \,g^{\ast}_{\pi\:in}(k)
       -\left(\del_{\la}\frac{g_{B\:in}(k)}{M}\right)
       \left(\del^{\la}\frac{g^{\ast}_{\pi\:in}(k)}{M}
       \right)\right\}\right]\nn\\
      & &-\sum_{l}M^2\left[\left(\del_{\mu}\frac{g_{\cb\:in}(l)}{M}\right)
       \left(\del_{\nu}\frac{g^{\ast}_{c\:in}(l)}{M}\right)
       +\left(\del_{\nu}\frac{g_{\cb\:in}(l)}{M}\right)
       \left(\del_{\mu}\frac{g^{\ast}_{c\:in}(l)}{M}\right)\right. \nn\\
      & &\qquad\quad+\left.g_{\mu\nu}\left\{\al\:
       g_{\cb\:in}(l)\,g^{\ast}_{c\:in}(l)
       -\left(\del_{\la}\frac{g_{\cb\:in}(l)}{M}\right)
       \left(\del^{\la}\frac{g^{\ast}_{c\:in}(l)}{M}\right)\right\}\right]\nn\\
      &=&0
    \eeqa
    by using (\ref{eqn:condition}) which is a necessary and sufficient
    condition for a BRST-invariant vacuum state.

  \subsection{The effective equation of $\phc$}
     The 1-loop effective equation of $\phc$ is
     given by
    \beqa
      \Box\phc&+&\left(M^2+\xi R\right)\phc+\frac{\la}2\phc^3
               -e^2\phc\langle U_{\mu}U^{\mu}\rangle
               +\frac32\la\phc\langle\rho^2\rangle\nn\\
              & +&\phc\left\langle\del_{\mu}\frac{\brB}{\phc}\del^{\mu}
                  \frac{\brB}{\phc}-2e^2\al\brB^2\right\rangle
                +2\phc\left\langle\del_{\mu}\frac{\brB}{\phc}\del^{\mu}
                \frac{\pi}{\phc}-2e^2\al\brB\pi
               \right\rangle\nn\\
              &+&2i\phc\left\langle\del_{\mu}\frac{\brcb}{\phc}\del^{\mu}
                \frac{\brc}{\phc}-2e^2\al\brcb\brc
                \right\rangle=0
    \eeqa
    This can be rewritten as follows.
    \beqa
      \Box\phc&+&\left(M^2+\xi R\right)\phc+\frac{\la}2\phc^3
                 -e^2\phc\langle U_{\mu}U^{\mu}\rangle
                 +\frac32\la\phc\langle\rho^2\rangle\nn\\
              &-&2i\phc\left\langle\left\{i\bolQB,\:\del_{\mu}
                 \frac{\brcb}{\phc}\del^{\mu}
                 \left(\frac{\pi}{\phc}+\frac12\frac{\brB}{\phc}\right)
                 -2e^2\al\brcb\left(\pi+\frac12\brB\right)
                 \right\}\right\rangle=0
    \eeqa
    This shows again that there is no backreaction from the BRST quartet
    for a physical state in spite of their condensation.

\setcounter{equation}{0}
\section{Summary and Conclusion\label{sec:sum}}
  We have investigated the U(1) Higgs model in spacetime-dependent background
  fields. In particular we choose a gauge fixing condition which is a
  generalization of the familiar $R_{\xi}$ gauge.

  The discrepancy between the in-vacuum and the out-vacuum leads to
  the condensation of physical and even unphysical particles. However, in case
  both vacuum states are physical, this can occur in a BRST-exact form.
  In other words, as is shown in \cite{hoho}, the BRST charge is invariant
  under the Bogoliubov transformation.
  The condensation of unphysical particles do not contribute to backreaction
  in the background field equations of a physical state
  because the corresponding terms are BRST-exact.

\appendix
\renewcommand{\theequation}
{A.\arabic{equation}}
\setcounter{equation}{0}
\section*{Appendix}
  \mbox{}

 1. Unitarity condition
  \beq
    \alU\alU^{\dag}-\beU\beU^{\dag}=\tilI,\qquad
    \alU^{T}\alU^{\ast}-\beU^{\dag}\beU=\tilI
  \eeq
  \beq
    \alU\beU^{T}-\beU\alU^{T}=0,\qquad
    \alU^{T}\beU^{\ast}-\beU^{\dag}\alU=0
  \eeq
  \beq
    \alBp\alBp^{\dag}-\beBp\beBp^{\dag}=I,\qquad
    \balBp^{T}\balBp^{\ast}-\bbeBp^{\dag}\bbeBp=I
  \eeq
  \beq
    \alBp\beBp^{T}-\beBp\alBp^{T}=0,\qquad
    \balBp^{T}\bbeBp^{\ast}-\bbeBp^{\dag}\balBp=0
  \eeq
  \beq
    \alBp\balBp^{\dag}-\beBp\bbeBp^{\dag}=I,\qquad
    \balBp^{T}\alBp^{\ast}-\bbeBp^{\dag}\beBp=I
  \eeq
  \beq
    \alBp\bbeBp^{T}-\beBp\balBp^{T}=0,\qquad
    \balBp^{T}\beBp^{\ast}-\bbeBp^{\dag}\alBp=0
  \eeq
  \beq
    \alcc\balcc^{\dag}-\becc\bbecc^{\dag}=I,\qquad
    \balcc^{T}\alcc^{\ast}-\bbecc^{\dag}\becc=I
  \eeq
  \beq
    \alcc\bbecc^{T}-\becc\balcc^{T}=0,\qquad
    \balcc^{T}\becc^{\ast}-\bbecc^{\dag}\alcc=0
  \eeq
  \beq
    \alr\alr^{\dag}-\ber\ber^{\dag}=I,\qquad
    \alr^{T}\alr^{\ast}-\ber^{\dag}\ber=I
  \eeq
  \beq
    \alr\ber^{T}-\ber\alr^{T}=0,\qquad
    \alr^{T}\ber^{\ast}-\ber^{\dag}\alr=0
  \eeq

  where

  \beq
    \tilI(ka,\kp\ap)=\de(ka,\kp\ap)
  \eeq
  \beq
    I(k,\kp)=\de(k,\kp)
  \eeq
  \beq
    (\al\al^{\dag})(k,\kp)=\sum_{\kpp}\:\al(k,\kpp)\al^{\ast}(\kp,\kpp),\quad
     \cdots
  \eeq

  \vspace{3ex}
  2.  Symmetry of $\la(k,\kp)$
  \beqa
    & &\sum_{\kpp}\left\{\al(k,\kpp)\be(\kp,\kpp)-\be(k,\kpp)\al(\kp,\kpp)
    \right\}=0\nn\\
    &\Longrightarrow&\sum_{\kpp}\left\{\al^{-1}(k,\kpp)\be(\kpp,\kp)
                     -\al^{-1}(\kp,\kpp)\be(\kpp,k)\right\}=0\nn\\
    &\Longrightarrow&\la(k,\kp)=\la(\kp,k)
  \eeqa

\vspace{0.5in}
\noindent{\em Acknowledgment.}

\vspace{0.2cm}
  The author deeply thanks Dr.~S.~Wada for his helpful comments and

\end{document}